\documentclass[prb,twocolumn,showpacs,amsmath,amssymb,superscriptaddress]{revtex4-1}
\usepackage{graphicx}
\usepackage{bm}
\usepackage{bbm}

\usepackage{subfigure}
\usepackage{amssymb}

\begin{document}
\title{\textbf{Disorder and localization effects on the  local spectroscopic and infrared-optical properties of   $\mbox{Ga}_{1-x}\mbox{Mn}_x\mbox{As}$}}

\author{Huawei Gao}
\affiliation{Institut f\"ur Physik, Johannes Gutenberg Universit\"at Mainz, 55128 Mainz, Germany}
\affiliation{Department of Physics, Texas A\&M University, College Station, Texas 77843-4242, USA}
\author{Cristian Cernov}
\affiliation{Department of Physics, Texas A\&M University, College Station, Texas 77843-4242, USA}
\author{T. Jungwirth}
\affiliation{Institute of Physics ASCR, v.v.i., Cukrovarnick\'a 10, 162 53 Praha 6, Czech Republic}
\affiliation{School of Physics and Astronomy, University of Nottingham, Nottingham NG7 2RD, United Kingdom}
\author{Jairo Sinova}
\affiliation{Institut f\"ur Physik, Johannes Gutenberg Universit\"at Mainz, 55128 Mainz, Germany}
\affiliation{Institute of Physics ASCR, v.v.i., Cukrovarnick\'a 10, 162 53 Praha 6, Czech Republic}
\affiliation{Department of Physics, Texas A\&M University, College Station, Texas 77843-4242, USA}

\date{\today}

\begin{abstract}
We study numerically the influence of disorder and localization effects on the local spectroscopic characteristics and infrared optical properties 
of $\mbox{Ga}_{1-x}\mbox{Mn}_x\mbox{As}$.  
We  treat the band structure and disorder effects at an equal level
by using exact diagonalization supercell simulation method. This method accurately describes the low doping limit and gives a clear picture of the transition to higher dopings, which captures the localization effects inaccessible to other theoretical methods commonly used. Our simulations capture the rich mid-gap localized states observed in  scanning tunneling microscopy studies and reproduce the observed features of the infrared optical absorption experiments. We show clear evidence of a disordered valence band model for metallic samples in which (i) there is no impurity band detached from the valence band, (ii) the disorder tends to localize and pull states near the top of the valence band into the gap region, and (iii) the Fermi energy is located deep in the delocalized region away from the mobility edge. We identify localized states deep in the gap region by visualizing the probability distribution of the quasiparticles and connecting it to their respective participation ratios. The analysis of the infrared-optical absorption data indicates that it does not have a direct relation to the nature of the states at the Fermi energy. 
\end{abstract}
\maketitle


\section{Introduction}
$\mbox{Ga}_{1-x}\mbox{Mn}_x\mbox{As}$ is a material system prototype, which incorporates carrier-mediated ferromagnetism into semiconductors.\cite{ohno1,ohno2,jungwirth_rmp,jungwirth_rmp14,Ohno_rmp14} Under equilibrium growth conditions, the solubility of Mn in GaAs is limited to $\sim 0.1\%$. Through non-equilibrium low-temperature molecular-beam-epitaxy (LT-MBE) technique, high Mn-doped samples with more than $1\%$ Mn can be obtained.\cite{singley,singley2} As has been shown in a recent systematic study,\cite{t2010,nc_nemec} high quality homogenous samples with reproducible characteristics can be prepared by introducing optimization of growth and post-growth annealing procedures for each doping concentration. Although its potential for applications has been curtailed due to growth limitations to achieve magnetic transition at room temperature, $\mbox{Ga}_{1-x}\mbox{Mn}_x\mbox{As}$ is a promising material for testing known magneto-transport, magneto-optical mechanisms and discovering of new magnetic phenomena.\cite{jungwirth_rmp,jungwirth_rmp14} 

However, even after many recent studies on this material, a common debate still exists regarding the electronic states near the top of the valence band and at the Fermi energy.\cite{jungwirth2} At small Mn doping, $x \lesssim 0.1\%$, $\mbox{Ga}_{1-x}\mbox{Mn}_x\mbox{As}$ is insulating and paramagnetic. The bound hole states introduced by Mn form an impurity band (IB) {\it detached} from the top of the GaAs valance band (VB). For higher Mn doping, $0.5\% \lesssim x \lesssim 1.5\%$, the isolated bound hole states start overlapping with each other and the IB starts mixing with the VB. The material is still insulating but ferromagnetism begins to occur.  At $x \sim 1.5\%$, the abrupt increase of low-temperature conductivity shows that the material becomes a degenerate semiconductor. In this metallic regime, one proposal states that the IB merges with the VB forming a disordered VB.\cite{jungwirth2} Another proposal assumes the Fermi level still resides in the narrow IB detached from the VB.\cite{singley2} However, the latter proposal, based on early spectroscopic studies on unoptimized samples,\cite{singley,singley2} is inconsistent with recent spectroscopic studies.\cite{nc_nemec} It has also been shown to be inconsistent with microscopic band structure theories.\cite{prl_masek} In this paper, through a more realistic treatment of disorder effects extrapolated from the small to large doping regimes, we show clear evidence of a disordered valence band model for metallic samples in which (i) there is no impurity band detached from the valence band, (ii) the disorder tends to localize and pull states near the top of the valence band into the gap region,  and (iii) the Fermi energy is located deep in the delocalized region away from the mobility edge.

Some early numerical studies either ignored disorder effects or treated them in the framework of mean field theory.\cite{timm,popescu,turek,alvarez,moca} In order to treat these effects more realistically, we use the exact diagonalization supercell method.\cite{yang} The advantage of this method is that the band structure and disorder effects are captured at an equal level which allows the study of disorder and localization effects on the local spectroscopic characteristics and infrared optical properties of $\mbox{Ga}_{1-x}\mbox{Mn}_x\mbox{As}$. The Hamiltonian includes the $\vec{k}\cdot\vec{p}$ description of the GaAs valance band, the Coulomb interactions at the Hartree level, short-range central-cell potential and kinetic exchange interaction. By diagonalizing this Hamiltonian numerically in the framework of the envelope function approximation,\cite{chow} eigenvalues and eigenfunctions can be obtained. Our wave function probability distribution visualization shows clear hydrogenic-like bound states for nearly isolated Mn impurities in low doping samples as well as the transition of these bound states to higher doping samples. The (local) density of states (LDoS/DoS) 
calculations capture the rich mid-gap localized states observed in  scanning tunneling microscopy studies.\cite{yazdani} 
The optical conductivity calculations reproduce the observed broad peak at $\sim 200$ meV of the infrared optical absorption measurements.\cite{singley,t2010} 
An analysis of the optical absorption data show no direct relation of this data to the nature of the states at the Fermi energy, contradicting the implied connection 
assumed by the IB models.   


The paper is organized as follows. Section II comprises a brief introduction of our simulation Hamiltonian. Section III provides the simulation results and discussions. In this section, we study the bound states properties in low Mn-doping limit first, then DoS and LDoS in the high concentration regime, followed by the analysis of the localized properties of the states, and finally the ac-conductivity calculations. Section IV presents the conclusions.

\section{Model Hamiltonian}
We use a phenomenological model employed by Yang {\it et al}.\cite{yang} The host semiconductor $\mbox{GaAs}$ valance band is described by the six-band $\vec{k}\cdot\vec{p}$ Kohn-Luttinger model.\cite{luttinger} Within the single particle approximation, the hole carriers interact with randomly placed Mn local moments via the Coulomb and the exchange interaction. 
A short-range central-cell correction is also considered in order to capture the difference in the electronegativity of the Mn and host Ga atom.\cite{bhatt} In real samples, there are charge and magnetic moment compensations due to As anti-sites and Mn interstitials. In this paper, we consider only the charge compensation introduced by As anti-sites because the magnetic moment compensation can be reduced by annealing procedures. The total Hamiltonian is given by:
\begin{eqnarray}
\hat{H} & =& \hat{H}^L + \sum_{I=1}^{N_{Mn}} \vec{S}_I\cdot \vec{s} J(\vec{r}-\vec{R}_I) \nonumber \\
&&+\: \sum_{I=1}^{N_{Mn}} \left(-\frac{e^2}{\epsilon |\vec{r}-\vec{R}_{{\rm Mn},I}|} -V_0 e^{-|\vec{r}-\vec{R}_{{\rm Mn},I}|^2/r_0^2} \right)\hat{I}\nonumber \\
&&+\: \sum_{K=1}^{N_{As}} \frac{2e^2}{\epsilon |\vec{r}-\vec{R}_{{\rm As},K}|}\hat{I},\label{ham}
\end{eqnarray}
where $\hat{H}^L$ describes the host valance band, $V_0$ term is the central-cell correction, $J(\vec{r}) = J_{pd}/[(2\pi a_0^2)^{3/2}]e^{-r^2/2a_0^2}$, $\hat{I}$ is a $6\times6$ unit matrix, and $\vec{s}=(\hat{s}_x,\hat{s}_y,\hat{s}_z)$ where $\hat{s}_{x,y,z}$ are $6\times6$ matrices which describe hole spins.\cite{abolfath} $\vec{R}_{{\rm Mn},I}$ and $\vec{R}_{{\rm As},K}$ are the positions of Mn and As respectively. The number of holes is given by relation $N_h = N_{Mn} - 2N_{As}$.

Our numerical method diagonalizes the single particle Hamiltonian exactly within a finite size cubic supercell with periodic boundary conditions.
In the supercell, Mn and As anti-sites are randomly distributed within the lattice and Mn spins are described by the classical $5/2$ local moment which are aligned in the $z$ direction at zero temperature. The phenomenological parameters we use are the same as the parameters in Ref. \onlinecite{yang}. We also treat the mutual interaction between holes by finding the self-consistent solution of the Hartree potential. We use 6 x 6 x 6 nm cube throughout our simulation. 
\begin{figure}[htbp]
   \includegraphics[width=0.8\columnwidth]{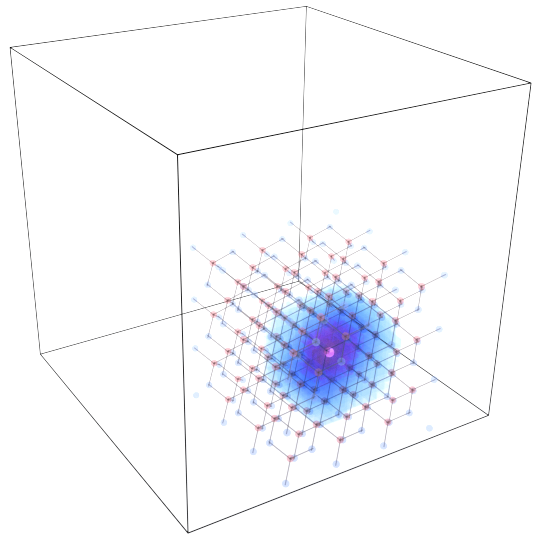} 
   \caption{(Color online) Wave function probability distribution of a single hole with a single Mn impurity in our supercell (black cubic outline). The lattice structure shown depicts the actual lattice constant with respect to the size of the supercell. The light color sphere in the center of the cloud is the Mn impurity.}
   \label{wav_1}
\end{figure}

\begin{figure*}[htbp]
\subfigure[]{\includegraphics[width=.5\columnwidth]{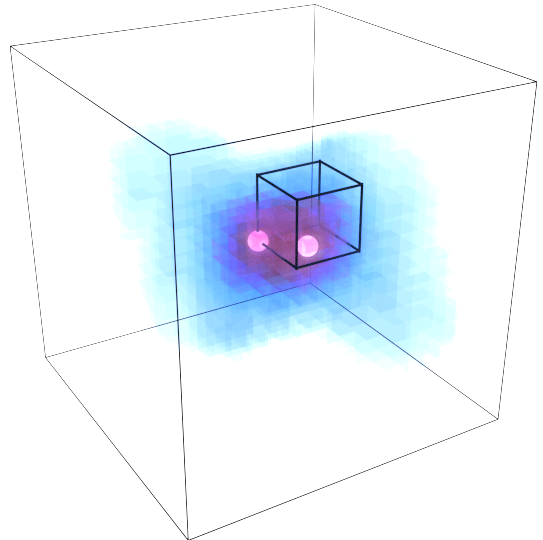} }
\subfigure[]{ \includegraphics[width=.5\columnwidth]{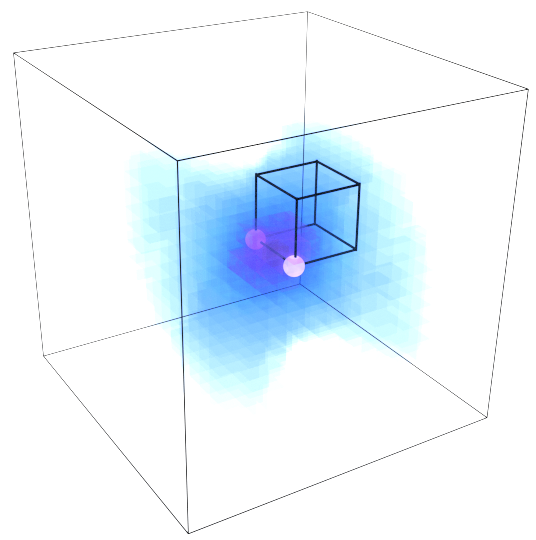}} 
\subfigure[]{ \includegraphics[width=.5\columnwidth]{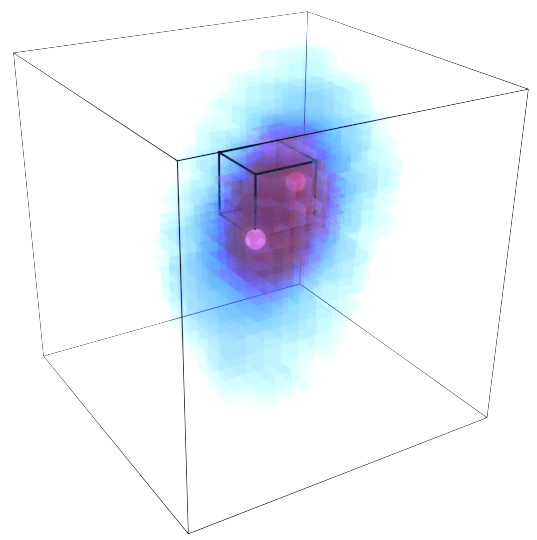}} 
\subfigure[]{ \includegraphics[width=.5\columnwidth]{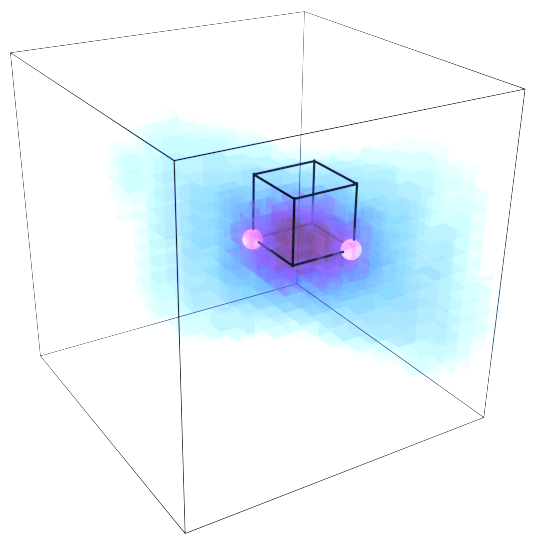}} 
\caption{(Color online) Wave function probability distribution for quasiparticle state with two Mn impurities sitting in the first to fourth nearest neighbor with respect to each other. The outer boxes are not the exact size of our simulation super cells. Light color spheres are the Mn and the small inner cubics show the actual size of unit cell of GaAs. (a) First nearest neighbor Mn with binding energy 276 meV, (b) Second nearest neighbor Mn with binding energy 232 meV, (c) Third nearest neighbor Mn with binding energy 222 meV, (d) Fourth nearest neighbor Mn with binding energy 199 meV.}
\label{wav_nn}
\end{figure*}

\section{Results and discussions}
\subsection{Low doping limit: \\bound holes \& pair-bonding states}
The advantage of our method is the ability to capture the low Mn doping as well as high Mn doping limit. In this section, we first study the low limit where there is only a single Mn impurity in our simulation supercell ($0.02\%$ Mn). Figure \ref{wav_1} clearly shows a hydrogenic-like bound state with a Bohr radius of $\sim$ 4 a$_{lc}$, where a$_{lc} = 0.565$ nm is the lattice constant of GaAs. The outer box shows our simulation supercell, and the lattice structure inside follows the actual lattice constant with respect to the box. The position of the Mn, shown as the light sphere in the middle of the distribution, is randomly chosen. The binding energy of this bound state is $\sim$ 40 meV, which is about $1/3$ of the experimental value for isolated Mn. The results  approach the experimental value with an increase in size of our simulation cell and the number of $k$ points for the envelope function expansion.

If two bound states are placed close to one another, one expects that bonding and anti-bonding states will be formed due to the interaction between the states. 
Depending on the strength of interaction due to this overlap, the energy difference can be very large. In our case, the bonding state will move deeper in the gap while the anti-bonding state will move to the valance band region. Figure \ref{wav_nn} shows our simulation results for the bonding state of two Mn impurities. Arranged from first to fourth nearest neighbor positions, the binding energies are 276 meV, 232 meV, 222 meV, and 199 meV respectively. To provide a better visualization, the images show a zoomed-in view of the probability distribution function, i.e.,  the outer box outline is not the actual size of our simulation; however, the inner cubic outline is the actual size of the lattice structure with respect to the probability distribution. The highly anisotropic structure of the probability distribution is consistent with STM experiments.\cite{stm} 
As expected, the bonding states have much larger binding energies than the isolated bound states.

\subsection{High doping regime}

\subsubsection{DoS and LDoS}

The impurity band model assumes that a detached band still persists in the high doping limit. In this section, we evaluate this high doping regime by studying the density of states (DoS) and local density of states (LDoS) in order to confirm whether or not the detached impurity band persists. As we know from the previous discussion, two nearby bound states will form bonding and anti-bonding states. In the high doping limit, there are many impurity pairs which give rise to a great amount of bonding states in the gap region. On the other hand, the anti-bonding states will contribute a large number of states to the valance band region. 

\begin{figure}[htbp]
 

 \includegraphics[width=0.9\columnwidth]{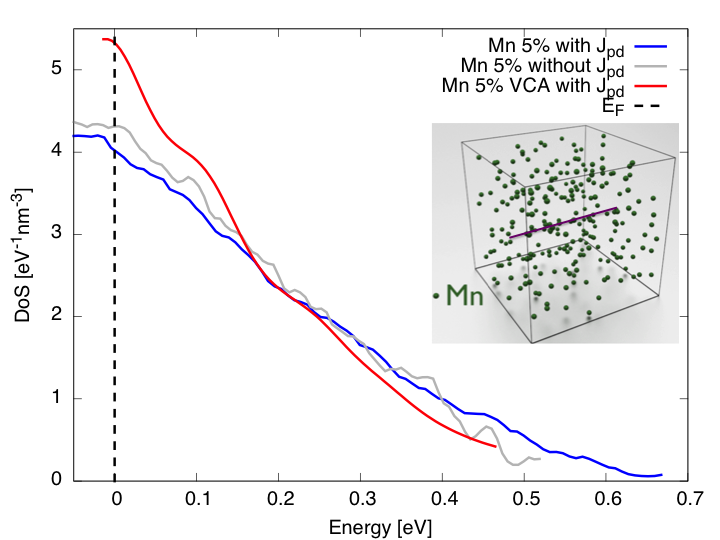}
 
 \caption{(Color online) Density of States (DoS) with and without exchange interactions for 5\% Mn doping without As compensations. We shift all the curves to make the Fermi energies to be zero as the vertical line shows. The red curve is the DoS obtained from VCA with an energy level broadening of 60 meV. The inset shows one typical Mn distribution in our simulation cell. The purple line in the cell shows the coordinate of the LDoS calculation (shown in Fig. 4).}
   \label{dos}
\end{figure}

The total DoS can capture the distribution of these states. Figure \ref{dos} shows the DoS plots for 5\% Mn doping, without As anti-site compensation. This figure shows the DoS with and without exchange interactions, i.e. ferromagnetic and non-ferromagnetic phases, as well as the DoS obtained from virtual crystal approximation (VCA).\cite{jairo} VCA assumes the wave vectors $k$ remain good quantum numbers with disorder treated as an energy spectrum broadening. As shown in the DoS plot, the VCA can only change the structure of the valance band  by slightly shifting the energy levels into the gap, whereas our more realistic treatment of disorder tends to pull states near the top of valance band deeper into the gap region. 
The kinetic-exchange interaction further splits energy levels and redistributes the states in the gap. As expected, there is an extensive weight of the DoS in the mid-gap region which arises from the deep bonding states created by neighboring impurities. However, these DoS curves show only a disordered valance band as opposed to an impurity band detached from the top of the valence band. 

\begin{figure}[htbp]
   \includegraphics[width=0.9\columnwidth]{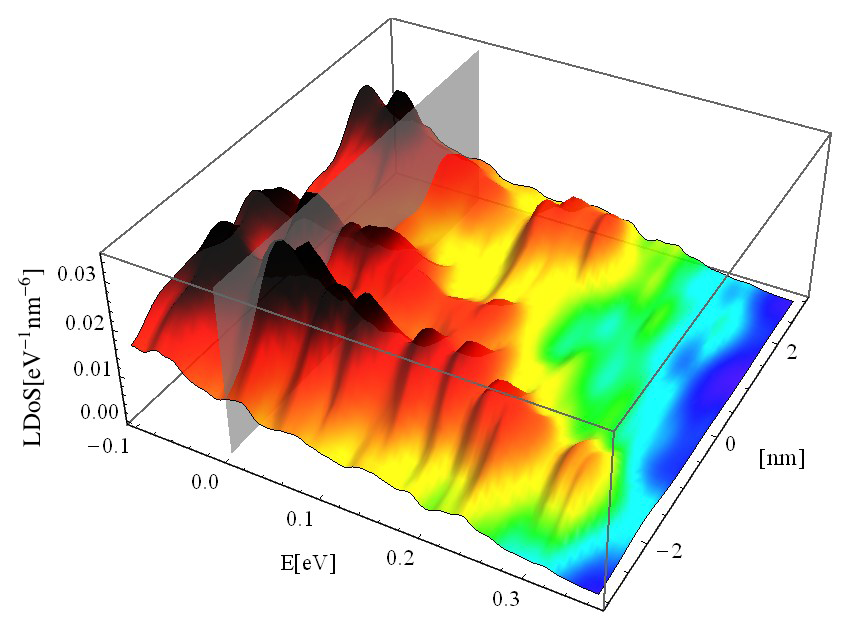} 
    \vspace{-0.3 cm}
   \caption{(Color online) Local Density of States (LDoS) for 5\% Mn density without As compensations. This plot shows different positions along the x-axis of our simulation supercell which is shown in Fig. \ref{dos}. The vertical plane shows the position of the Fermi energy. The results agree qualitatively with the experiments in Ref.~\onlinecite{yazdani}.}
   \label{ldos}
 \vspace{-0.6 cm}
\end{figure}

To further study the spatial inhomogeneity of the distribution of states in the gap region, we calculate the LDoS which can be qualitatively compared with the STM experiments.
 Figure \ref{ldos} shows our simulation results along a line in our simulation cell. The gray vertical plane is the Fermi energy. The peaks formed by bonding states in the gap region are consistent with the STM measuments,\cite{yazdani}  which indicates our model captures the disorder effects correctly. 
 We emphasize here that in our approximation, which ignores the Fock-exchange contribution, we cannot capture the expected reduction of the DoS at the Fermi energy
 in disorder systems. 

%
%

\subsubsection{Localization properties of the states in the high doping regime}

\begin{figure*}[htbp]
\subfigure[]{  \includegraphics[width=0.8\columnwidth]{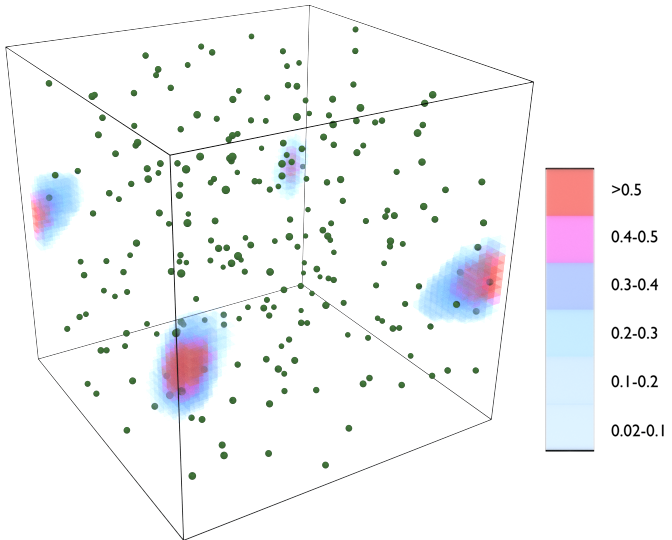}}
\subfigure[]{  \includegraphics[width=0.8\columnwidth]{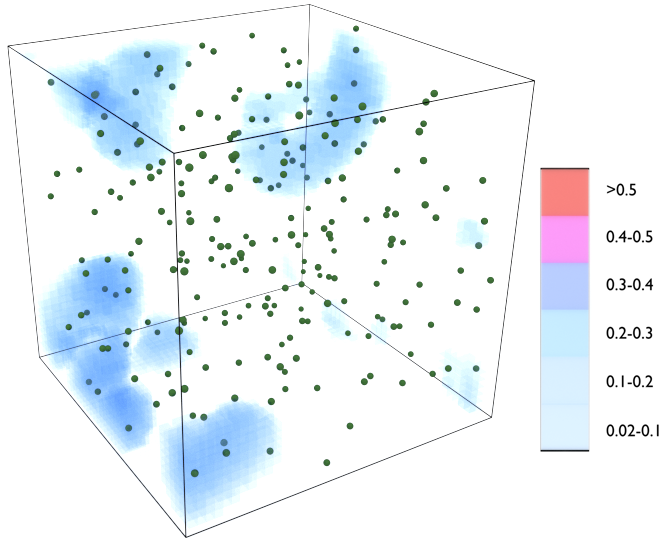}}
\\
\subfigure[]{  \includegraphics[width=0.8\columnwidth]{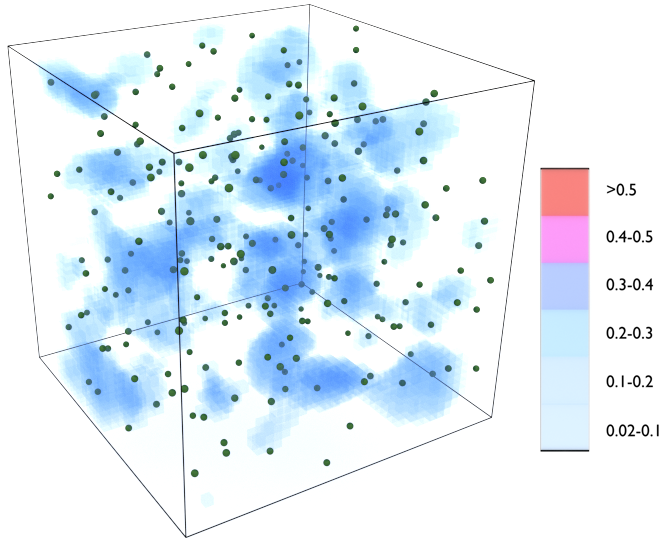}} 
 \subfigure[]{ \includegraphics[width=0.8\columnwidth]{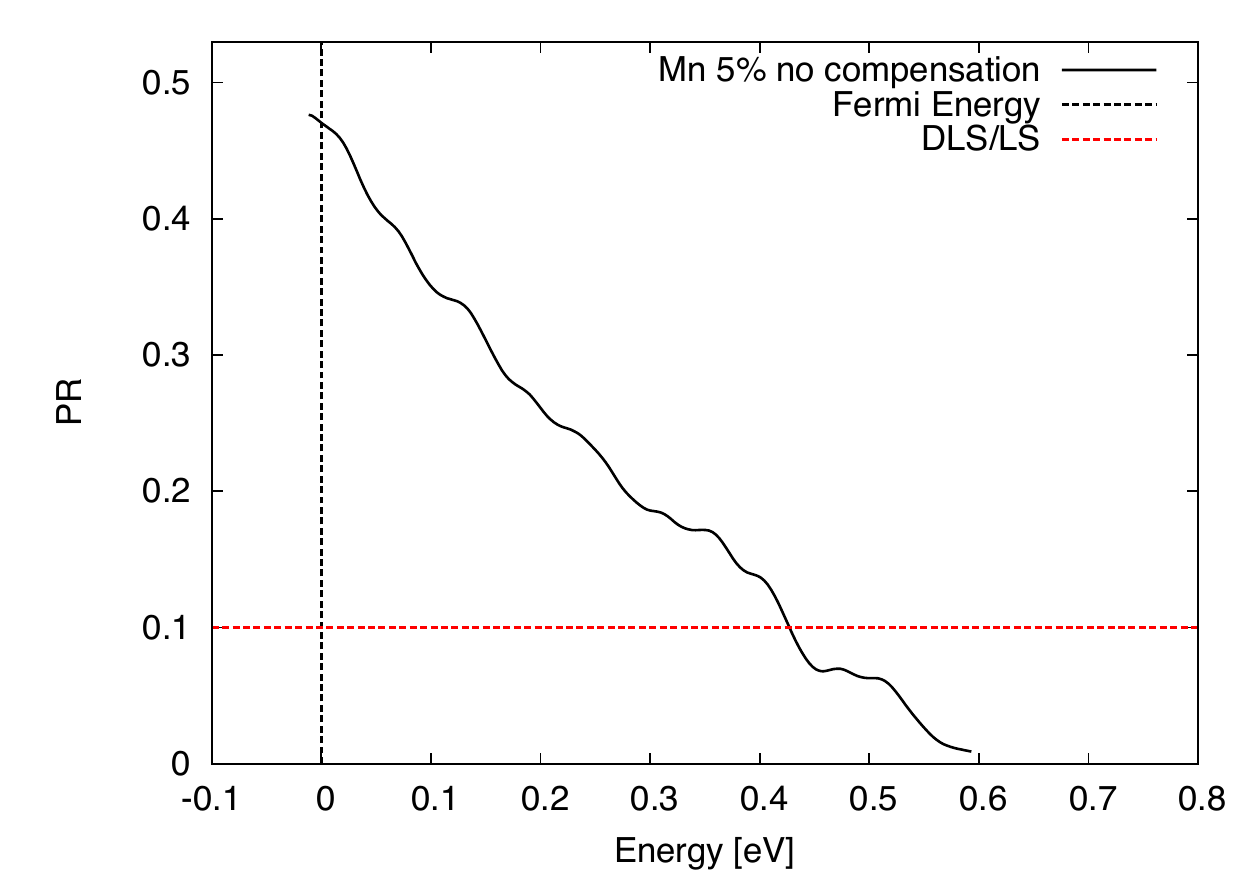}} 
 \caption{(Color online) (a) Wave function probability distribution with PR = 0.015, (b) Wave function probability distribution with PR = 0.1, (c) Wave function probability distribution with PR = 0.2, (green spheres denotes Mn impurities), (d) Participation ratios.}
 \label{pr}
\end{figure*}

From our DoS study, an impurity band detached from the valence band does not persist for high Mn doping. To further confirm this observation, we study the localization character of the states throughout the spectrum. If an impurity band were to exist in the high doping limit, we would expect localized states in both tails of the impurity band. As we have shown, bound states in the low Mn doping limit ($< 0.1 \%$) are well localized indicating the existence of an impurity band for this limit. The probability distribution of these wave functions are concentrated in a small region with a size of several GaAs lattice constants. 
When increasing Mn doping, some of the hole states which are well isolated from others will stay localized, but for some others their wave functions begin to overlap with each other leading to delocalization. 
To have a quantitative description of localized and delocalized states, we use participation ratios (PRs) to characterize the extent of localization  of states and compare with the probability distributions directly to establish the criteria for localized states. Then, we group these states as localized and delocalized based on this criteria. The results show that our model is able to capture the localization physics with its more realistic treatment of disorder. PR is defined as the inverse of the integral over the simulation volume of the wave function to the fourth power.\cite{prs,pr}
\begin{equation}
\mbox{PR} = \frac{1}{V\int_V d^3r |\Psi(\vec{r})|^4},
\end{equation}
where V is the volume of the simulation cell. $\Psi(\vec{r})$ is the normalized wave function. A simple example gives the meaning of the PR. For delocalized states, $|\Psi(\vec{r})|^2\sim 1/{V}$, so PR $\sim1$, but for localized states, the wave function probability is concentrated in an area much smaller than V. The PR has the order of $1/V$, which is much smaller than 1 if V is large enough. Another way of distinguishing localized and delocalized states is using size scaling of the PRs. With this method, the PR is almost constant for delocalized states, but scales as $1/V$ for localized states. The disadvantage of using size scaling of PR is the computation time. In this paper, 
similarly to other studies in disorder systems,
we use the magnitude of PR and relate it to the probability distribution for hole state wave functions to distinguish localized and delocalized states. 

\begin{figure}[htbp]
   \centering
   \includegraphics[width=0.9\columnwidth]{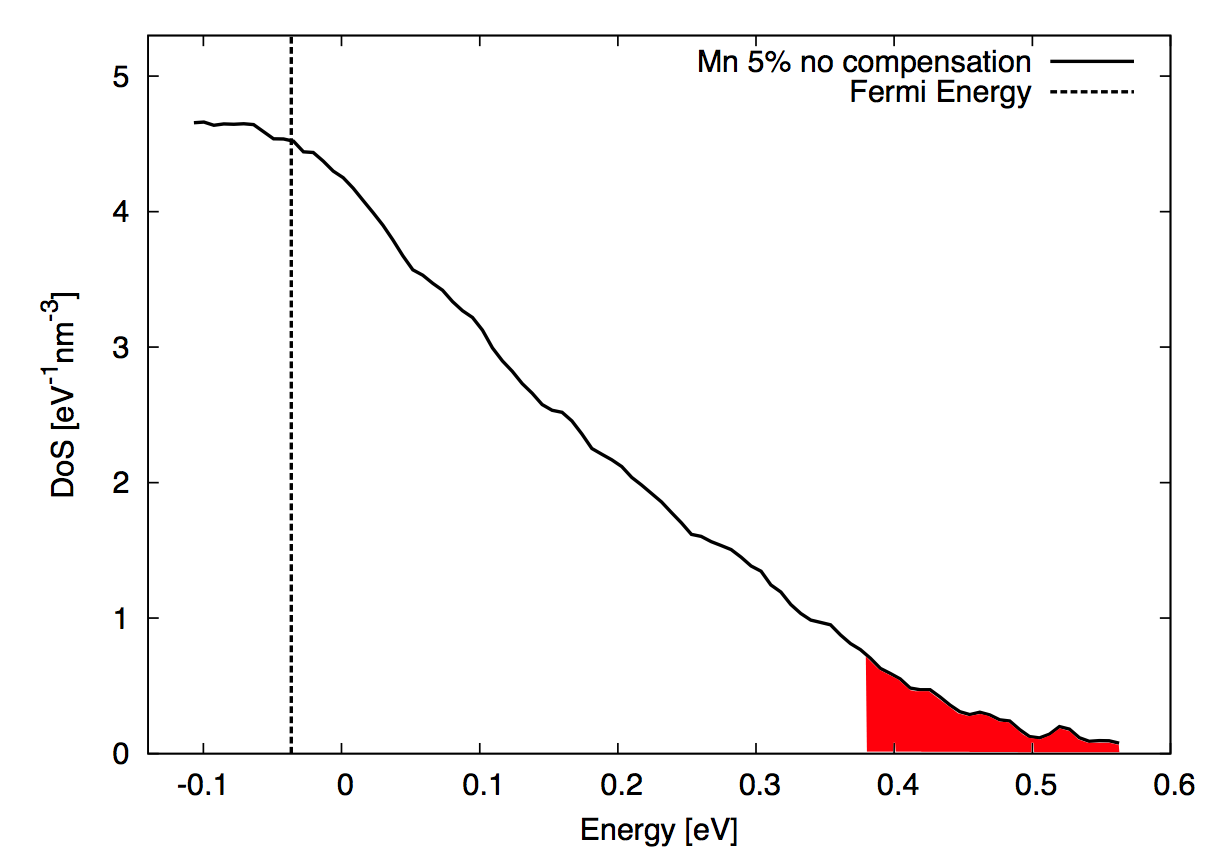} 
   \caption{(Color online) Density of States (DoS) for 5\% Mn without As compensations. Red area shows the localized states}
   \label{dos_c}
\end{figure}

\begin{figure}[htbp]
   \centering
   \includegraphics[width=0.9\columnwidth]{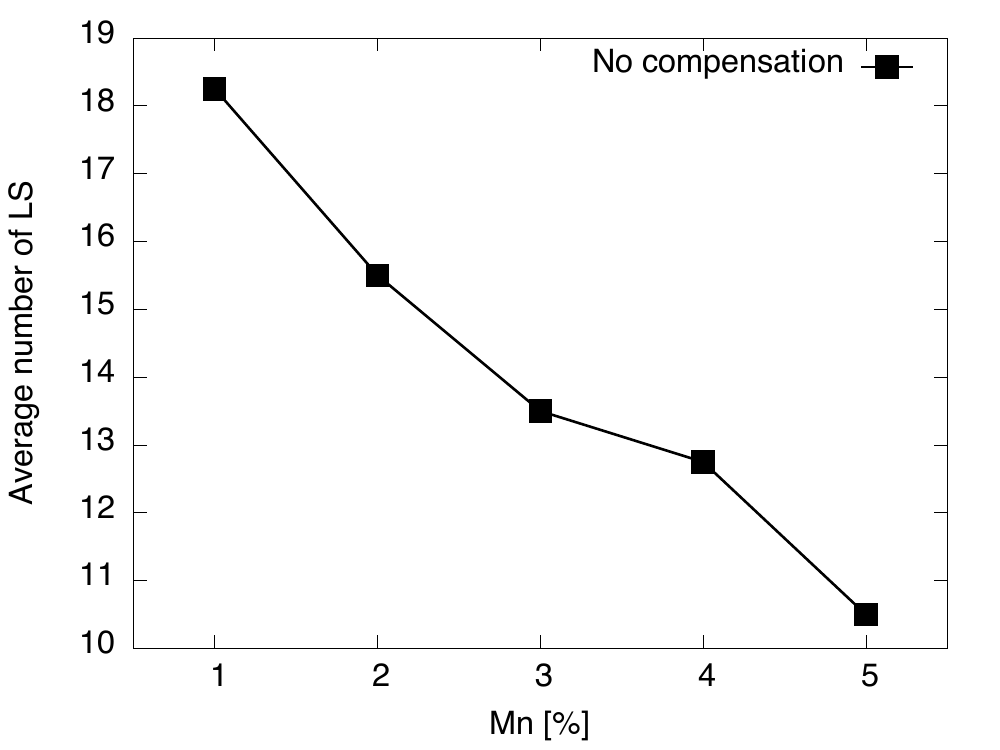} 
   \caption{Average number of localized states in our simulation cell. Data obtained by averaging over 4 disorder realizations.}
   \label{fig_mn}
\end{figure}

Figure \ref{pr} shows the participation ratios as a function of energy and the typical wave function probability distributions with different participation ratios. For our choice of V, states with PR $<$ 0.1 are localized states with their wave function concentrated in a small area. For PR $>$ 0.1 but close to 0.1, these states are a transition  from localized to delocalized states which define the mobility edge. Within this definition there is of course no sharp separation of localized and delocalized states. 
This would require the use of the finite-size scaling method. However, for the purposes of our study it suffices to treat states with PR$<0.1$ as localized and above it as delocalized. 
With this criteria, the red region in Fig.~\ref{dos_c} shows the area of the localized states in the DoS plot, meaning that states deep in the gap are well localized which is consistent with our previous discussion. These states are bonding states formed by nearby Mn impurity pairs with strong interactions. However, if more Mn are doped into this system, localized holes have more places to hop around becoming delocalized due to the increasing number of nearby Mn. There will be less localized states for higher Mn doping samples. Figure \ref{fig_mn} shows the simulated results for different Mn doping without compensation obtained by averaging over four disorder realizations. It is clear that the number of localized states decreases with increasing Mn doping.  
In our results only states deep in the gap region are localized far away from the Fermi energy, which does not support the impurity band picture.

\subsection{Ac-conductivity}

\begin{figure}[htbp]
   \centering
   \includegraphics[width=0.9\columnwidth]{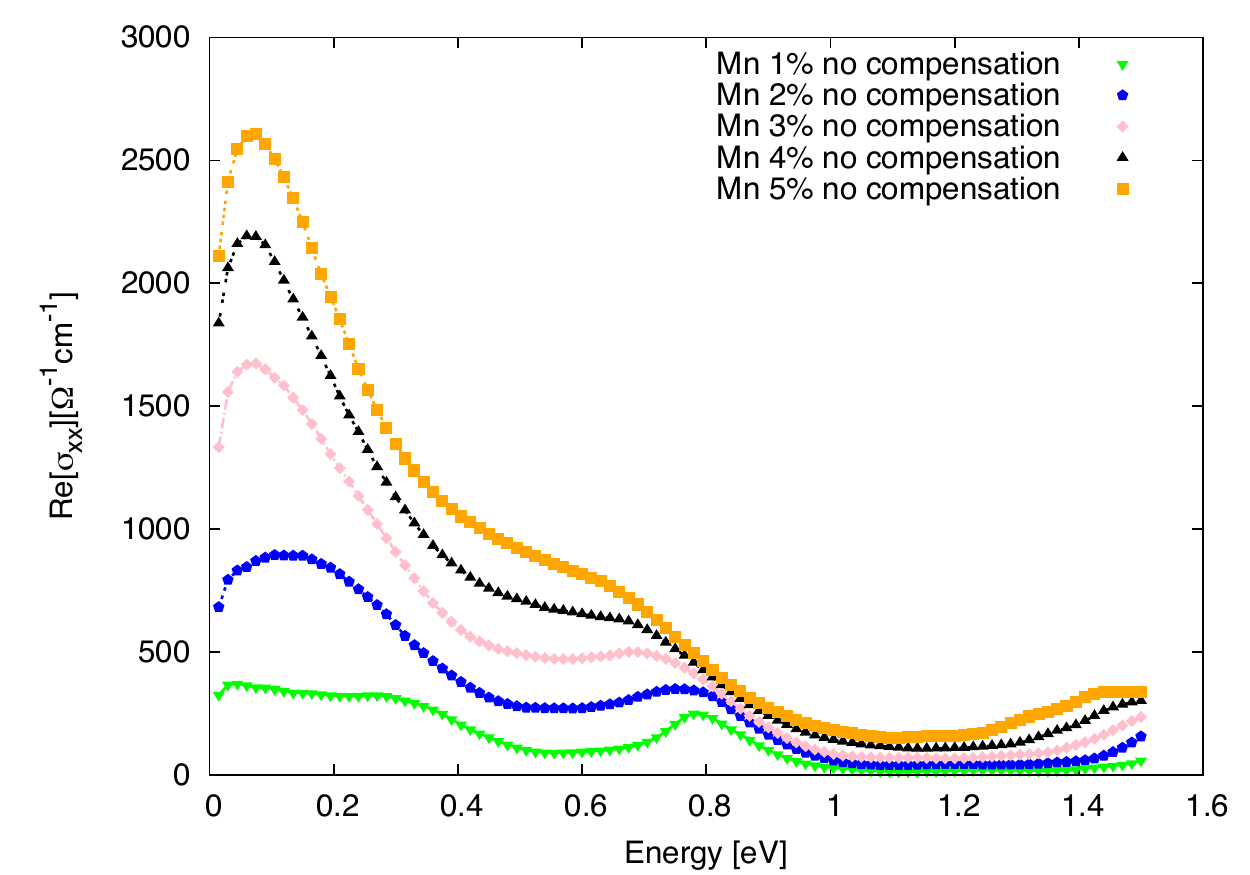} 
   \caption{(Color online)Real part of $\sigma_{xx}$ for various Mn density without As compensations. Data obtained by averaging over 4 disorder realizations.}
   \label{fig_act}
\end{figure}

\begin{figure}[htbp]
   \centering
   \includegraphics[width=0.9\columnwidth]{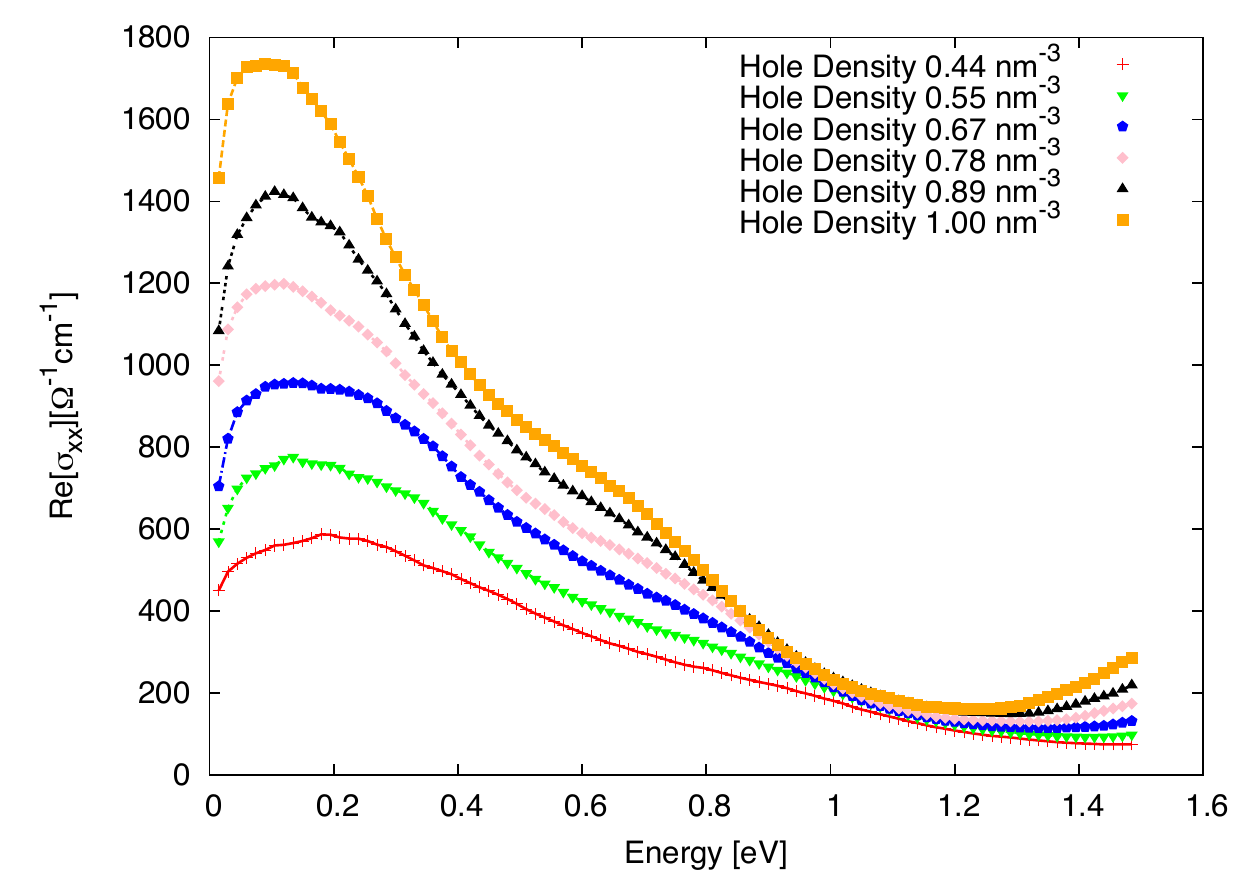} 
   \caption{(Color online) Real part of $\sigma_{xx}$ for 5\% Mn doping and different hole density due to As compensation. Data is obtained by averaging over 4 disorder realizations.}
   \label{fig_act_5}
\end{figure}

In this part, we will study the ac-conductivity to explain the spectroscopic experimental observation which was used to support the impurity band model. The method used here is the standard linear response Kubo formalism. Early experiments of infrared ac-conductivity measurements on $\mbox{Ga}_{1-x}\mbox{Mn}_x\mbox{As}$ shows a broad peak near 200 meV for metallic samples. A red shift was observed when increasing hole concentration, which is the key evidence of impurity band model.\cite{singley2} Later on, similar experiments on optimized growth and post-growth annealing samples show a blue shift when increasing hole concentration.\cite{t2010} By post-growth annealing procedures, the density of compensating defects and other unintentional impurities can be greatly reduced. To compare with these experiments, we will study uncompensated samples first. Figure \ref{fig_act} shows the ac-conductivity for Mn dopings ranging from 1\% to 5\% without compensation. For each doping rate, we do see a broad peak in the low energy region $\sim$100 meV. When increasing Mn doping, there is a red shift when Mn is less than $3\%$ and no shift for higher dopings. We expect that compensation plays an important role. Aside from the peak $\sim$100 meV, there is one other peak $\sim$800 meV for lower Mn doping samples, which arises 
 from an unphysical finite size effect and band cut-off for the purposes of the calculation. To study the effects of compensation, we fix the Mn doping to be $5\%$ and consider As anti-sites as a source of hole compensation. Figure \ref{fig_act_5} shows our simulated results for different hole densities. Compared to the uncompensated samples, the peaks are now shifted from $\sim$100 meV to $\sim$200 meV. There is still no prevailing red shift with increasing hole density.

The simulated results show our model is able to capture the mid-infrared peak in optical absorption measurements by considering the transitions from states below the Fermi energy to states above the Fermi energy and deep in the gap region. The Fermi energy resides in the delocalized disordered GaAs valance band region.
This demonstrates that the assumption of the detached IB model is not valid from its starting point in such metallic samples. As shown in the ac-conductivity simulation, the peaks for compensated samples move towards high energy compared to uncompensated samples. Hence one can get any pattern of the peak shifts 
if less and more compensated (annealed and unannealed) samples are mixed.
Experimentally, the actual density of compensation defects is hard to determine. Aside from As anti-sites, there are other types of charge and moment compensations. So, based on our simulations, the peak experimentally observed  shifts of the  $\sim$200 meV cannot be used to make any direct conclusions 
regarding the nature of states at the Fermi energy.

\section{Conclusions}

In this paper, we provide a numerical study of the effect of disorder and localization on the states near the top of the valence band and in the gap region of Mn-doped GaAs. Through exact diagonalization of our model Hamiltonian, the low and high doping limits are captured. The local spectroscopic properties are studied and compared with STM experiments by simulating the DoS and LDoS. Furthermore, via visualization of the probability distribution of the quasiparticles and the connection to their respective participation ratios, we separate the states into localized and delocalized. It is shown for metallic samples that the states deep in the energy gap are well localized and states near the Fermi energy are delocalized. The optical conductivity with different Mn dopings and As compensations are then simulated. Our results show that the peak at $\sim$200 meV is due to the transition of states below Fermi energy to states in the gap region.  There is no direct relation between the optical absorption measurements and the nature of states at the Fermi energy. Our overall results are consistent with the disordered valence band model and 
disagree with the assumption of the IB models of a detached impurity band from the valence band. 

\begin{acknowledgments}
The authors thank Vivek Amin and Erin K. Vehstedt for helpful discussions and suggestions on data visualization. Part of our computer simulation was carried out in the supercomputer in FZU in CZ and University of Hamburg in Germany. This work was supported by the US grants NSF-DMR-1105512, ONR-n000141110780, and by the Alexander Von Humboldt Foundation.
We also acknowledge support from the EU European Research Council (ERC) Advanced Grant No. 268066, the Ministry of Education of the Czech Republic Grant No.LM2011026, and the Grant Agency of the Czech Republic Grant No. 14-37427G.
\end{acknowledgments}



\end{document}